\documentclass[12pt]{IEEEtran}
\usepackage{amssymb}
\usepackage{float}
\usepackage{amsmath}
\usepackage{graphicx}
\def\sv{\left [ \begin{array}{c}}
\def\ev{\end{array} \right ]}
\def\smtwo{\left [ \begin{array}{cc}}
\def\bfs{\mathbf{s}}
\def\bfp{\mathbf{p}}
\def\bfq{\mathbf{q}}
\def\bfx{\mathbf{x}}
\def\bfr{\mathbf{r}}
\def\bfy{\mathbf{y}}

\def\bfe{\mathbf{e}}
\def\bfb{\mathbf{b}}
\def\bfPhi{\mathbf{\Phi}}
\def\em{\end{array} \right ]}
\def\ploss{n}
\def\emlocs{N}
\def\sensors{M}

\begin{document}

\title{Geolocation of Multiple Noncooperative Emitters Using Received Signal Strength: Sparsity, Resolution, and Detectability}

\author{Kurt~Bryan,~\IEEEmembership{Member,~IEEE,}
        Deborah Walter,~\IEEEmembership{Member,~IEEE,}
\thanks{Kurt Bryan is with the Department
of Mathematics, Rose-Hulman Institute of Technology, Terre Haute,
IN, 47803 USA e-mail: kurt.bryan@rose-hulman.edu.}
\thanks{Deborah Walter is with the Department
of Electrical and Computer Engineering, Rose-Hulman Institute of Technology, Terre Haute,
IN, 47803 USA e-mail: deborah.walter@rose-hulman.edu.}
\thanks{This work was supported in part by the Air Force Office of Scientific Research (AFOSR), FA9550-15-F-0001.}
\thanks{Copyright 2019 IEEE.  Personal use of this material is permitted.  Permission from IEEE must be obtained for all other uses, in any current or future media, including reprinting/republishing this material for advertising or promotional purposes, creating new collective works, for resale or redistribution to servers or lists, or reuse of any copyrighted component of this work in other works.}}

\maketitle

\begin{abstract}

In this paper we investigate the problem of locating multiple non-cooperative radio frequency (RF) emitters using only received signal strength (RSS) data.
We assume that the number of emitters is unknown and that individual emitters cannot be distinguished in the RSS data. Moreover, we assume that the environment in which
the data has been collected has not been mapped or ``fingerprinted'' by the prior collection of RSS data. Our primary interest is the
limiting resolution that can be obtained by this type of data, and the lowest power emitters that can be detected, as a function of noise
level, sensor geometry, and other variables. We formulate the recovery problem as one of
sparse approximation or compressed sensing, and investigate an appropriate recovery algorithm for this setting, and use it to illustrate our
conclusions. We also include a reconstruction based on sampled data we collected, to illustrate the reasonableness of our
parameter choices and conclusions.

\end{abstract}


\section{Introduction}
\label{sec:intro}

Locating radio frequency (RF) sources from remotely collected RF data is an essential task in many settings, and is commonly referred to as RF \textit{localization},
or \textit{geolocation}.
Applications are numerous, for example, the localization of subscribers in cell phone or other wireless networks (indoor or outdoor, see \cite{sayed,hadi,feng,weiss,Patwari}).
Localizing transmitters in a cognitive radio network (\cite{daponte,bazerque,li}) allows for the more efficient allocation of network resources, for example, frequency
bands. Autonomous vehicles may rely on RF localization to augment navigation \cite{daponte}. In military applications we may be tasked with geolocating RF transmitters
that are non-cooperative or evasive \cite{butler,king,daponte,Walt1,brywal}. See \cite{sayed} for a number of other applications.

A variety of techniques for localizing RF emitters from remote data have been developed. Some techniques use range information deduced from the signal time-of-arrival (TOA), time-difference-of-arrival (TDOA), or the received signal strength (RSS), perhaps collected from multiple sensors at spatially diverse locations. Others, such as angle-of-arrival (AOA), rely on directional information collected from sensors. We may or may not have information about the nature of the RF signals, e.g., emitted power or correlation of measured data from distinct sensors.  The accuracy of the resulting position estimates depends on uncertainties in the channel models, sensor placement, and precision of the data collected.

The accuracy of RSS as a method for geolocation is known to suffer from multipath and shadowing effects, but
RSS-based localization methods have the advantage that sensor design can be low-complexity; complicated timing, synchronization, or other sophisticated hardware is not needed.
Thus, the sensors can be relatively low-cost and low-power. The availability of such sensors is particularly important when many sensors are required, or the sensors are required to be battery-powered (e.g., remote or mobile sensors). Since Received Signal Strength Indicator (RSSI) values are available directly from systems implementing standard communication protocols (see for example, \cite{adams}), many WLAN applications do not need any additional hardware to implement RSS-based localization algorithms.

In this work we will focus on the problem of geolocating a ``small'' but unknown number of non-cooperative RF emitters using RSS measurements from multiple sensors dispersed geographically, or alternatively, from a single sensor on a moving platform, or some combination thereof. In particular,
we are interested in methods for estimating the best possible resolution one can obtain from RSS-based location estimates.

Localization of RF sources from RSS data has been considered before (\cite{weiss,whitehouse,fidan,hadi,deFreit}) in a variety of scenarios. Some (\cite{cevher,fengi,hadi,fann2}) have taken the rough approach we use---a compressed sensing view that exploits spatial sparsity in assuming a small number of emitters are present.
But many focus on situations
in which the emitters are cooperative \cite{whitehouse}, or only one emitter is present, or emitters can be distinguished in some manner (\cite{fidan,whiting,fengi}) in the data. Since RSS-based localization relies on a propagation model relating signal strength and distance to an emitter(s), RSS-based methods suffer if the signal strength model is inaccurate. Hence some prior work (\cite{hadi,cevher,tian,whitehouse}) assumes that the environment has been ``fingerprinted,'' that is, sensors have been placed in known locations (``anchors'', \cite{patwari2}), and then empirical measurements taken to map the RF environment. This improves the channel model and accuracy of emitter location estimates.
Some methods focus on prediction of lower bounds for the variance of location estimates from RSS data \cite{Koorapaty,Chitte1,Chitte2}.

We consider the problem of geolocating multiple non-cooperative RF emitters in an outdoor environment with a
known (at least approximately) channel model. We assume that RSS data is collected by multiple RF sensors (possibly mobile, e.g.,
mounted on UAVs) whose location(s) are known. We specifically focus, for illustrative purposes, on the case in which multiple elevated sensors are used in an unobstructed open-air
scenario, with stationary emitters on the ground, though our analysis is not tied to this arrangement. In particular, we assume that the emitters transmit at a common (known) frequency, such that:
\begin{itemize}
\item The number of emitters is not known, but is ``small,'' in a sense to be quantified later.
\item Emitter signals cannot be distinguished by any characteristic in the time or frequency domain. Thus the RSS data collected by any sensor is the ``aggregate'' power summed over all emitters.
\item The RF sensors are ``limited'' in number and have isotropic sensitivity, so no directional information is available.
\end{itemize}
From such data we seek to recover the number of emitters, the location of each, and possibly the power at which each emitter transmits.

The unique contributions outlined in this paper are to:
\begin{itemize}
\item Determine the limiting resolution (ability to distinguish two close emitters) from this type of data, as a function of the data noise/uncertainty level, sensor placement,
channel attenuation model, and other relevant physical parameters.
\item Determine the limiting power threshold for an emitter's ``detectability'' (the lowest power emitter than can be detected) as a function of the above-mentioned quantities.
\item Demonstrate that an appropriate algorithm that makes use of the above assumptions can in fact determine the number of emitters present and their locations.
\item Use this algorithm to illustrate our conclusions on resolution and emitter detectability.
\end{itemize}

In Sections \ref{sec:probform} and \ref{sec:sparserecov} below we formulate the problem of locating RF emitters from RSS data as one of finding
a sparse solution to an underdetermined linear system of equations, and include an appropriate noise model. We then examine the notion of \textit{coherence}, which
plays a central role in our analysis of resolution and emitter detectability, and then briefly examine an appropriate algorithm for solving the resulting
system. In Section \ref{sec:analysis} we analyze the resolution that can be obtained with this type of data, and the limits on emitter detectability. The algorithm
developed is used to illustrate our conclusions with computational examples. Finally, in Section \ref{sec:data} we detail
data we collected to validate our channel model parameters, and the geolocation of an emitter from measured RSS data.

\section{Problem Formulation}
\label{sec:probform}

\subsection{System Model}
\label{subsec:sysmodel}

Consider the problem of identifying an unknown
number of non-cooperative emitters located on the ground using RSS data from airborne sensors. Specifically, let $\Omega\subset\mathbb{R}^2$ and assume the emitters lie at
three-dimensional coordinates $(x,y,z)$ with $(x,y)\in\Omega$ and $z=0$.

Although a communication network may employ a number of different frequencies, it is often the case that those of interest for geolocation are relatively few. Several methods exist to detect and classify signals by their frequency content (see, for example, \cite{roy,schmidt}).  We will not focus on this aspect of the problem, but rather assume that a frequency (or narrow range of frequencies) of interest has been identified, and that the emitters of interest are operating at these
frequencies.

When the number of emitters is sufficiently small, localizing them is a problem well-suited to
formulation in the context of compressed sensing, that of finding a sparse solution to a linear system of equations, where ``sparse'' means that most components in the relevant solution vector are zero (or close to zero.)
Specifically, let $S=\cup_{i=1}^{\emlocs} \bfr_i$, where $\bfr_i=(x_i,y_i)$, be a subset of $\emlocs$ distinct points in $\Omega$; these will be the potential locations of any emitters. These points should be chosen to provide a reasonable sampling of the potential locations of any emitters. For example, if $\Omega$ is a rectangle it may be convenient to define $S$ as the nodes on a finely-spaced rectangular grid. It is not essential that emitters be located precisely at any of the $\bfr_i$.

Suppose there are $\sensors$ sensors that measure the RSS at known $(x,y)$ positions $\bfs_j = (a_j,b_j)$, each at a fixed altitude $h$ above the $xy$-plane (we take each at the same altitude only for simplicity; the sensors need not be at a single altitude nor directly above $\Omega$.)  The $\bfs_j$ may represent distinct sensors, or a single sensor taking data at different points along a path, or some combination thereof. We assume that the sensors' antennae are isotropic, though more complicated antenna patterns are easily accommodated in the analysis. The distance $r_{ij}$ from the $j$th sensor to the $i$th point in $S$ is $r_{ij} = \sqrt{\|\bfs_j-\bfr_i\|^2_2+h^2}$ where $\|\cdot\|_2$ denotes the usual Euclidean norm in the plane.

One common model for the power $P_{ij}$ received at sensor $j$ from an emitter at position $\bfr_i$ is that $P_{ij} = p_i (r_0/r_{ij})^{\ploss}$ where $p_i\geq 0$ is a reference power measured at distance $r_0$ from the emitter $i$ and $\ploss$ is the \textit{pathloss exponent} that governs the attenuation of the signal power as a function of distance; see
\cite{king,martin}. In the ideal case the RSS at sensor $j$ from all emitters is then modeled as
\begin{equation}
\label{attenmodel0}
d_j = \sum_{i=1}^{\emlocs} P_{ij} = \sum_{i=1}^{\emlocs} p_i \left (\frac{r_0}{r_{ij}}\right )^{\ploss}
\end{equation}
This assumes receiver antennas are equally sensitive, isotropic, and that the emitters are isotropic and incoherent. If no emitter is present at position $\bfr_i$ then $p_i=0$, so if few emitters are present we expect $p_i>0$ for only a few indices $i$.

We amalgamate the data $d_j$ into a column vector $\mathbf{d}_0\in\mathbb{R}^{\sensors}$ and express the ideal RSS data (\ref{attenmodel0}) in matrix form,
\begin{equation}
\label{attenmodel}
\mathbf{d}_0 = \bfPhi \bfp_0.
\end{equation}
Here $\bfPhi$ is the \textit{measurement matrix}, an $\sensors\times \emlocs$ matrix with known entry $(r_0/r_{ij})^{\ploss}$ in row $i$, column $j$. The vector $\bfp_0\in\mathbb{R}^{\emlocs}$ has $i$th entry $p_i$, the reference power of the emitter at $\bfr_i$, and is sparse if few emitters are present. Note that the entries of $\bfPhi$ are known. The $j$th row of $\bfPhi$ embodies the data from the sensor at position $\bfs_j$, and the $i$th column corresponds to a potential emitter location $\bfr_i$. We assume that we can measure the quantity $\mathbf{d}_0$, the power received by each sensor. The problem of interest is to recover an estimate of $\bfp_0$ from $\mathbf{d}_0$ and $\bfPhi$. Of course $\mathbf{d}_0$ will be corrupted by noise or other error.

\subsection{Measurement Noise Model}
\label{subsec:noisemodel}

Departure of measured RSS data from the ideal model above is consider at length in, for example, \cite{zanella}. We assume that data has been
suitably processed to eliminate the effects of so-called ``fast-fading'' and that the error that remains conforms to the standard log-normal
noise model. Specifically, if an emitter with reference power $p_i$ is present at location $\bfr_i$, the contribution to the data $d_j$ collected at the $j$th sensor
from this emitter is of the form
$$
d_j = p_i \left (\frac{r_0}{r_{ij}}\right )^{\ploss} e^{\eta R_i}
$$
where $\eta=\ln(10)/10$ and $R_i$ is a normal random variable with mean $0$ and standard deviation $\sigma_{dB}$. Note that $e^{\eta R_i} = 10^{R_i/10}$.
Here $\sigma_{dB}$ is the noise level in dB. Values for $\sigma_{dB}$ vary widely depending on the setting, but the application of interest here (outdoors, a relatively
open and obstruction-free area) values from $2$ to $5$ dB are common; see \cite{whitehouse} or our data in Section \ref{subsec:rssmeas}.

For multiple emitters we take
\begin{equation}
\label{noisemodel1}
d_j = \sum_{i=1}^{\emlocs} p_i \left (\frac{r_0}{r_{ij}}\right )^{\ploss} e^{\eta R_{ij}}
\end{equation}
with the additional assumption that the $R_{ij}$ are independent. The model in (\ref{noisemodel1}) is valid when the sensors are sufficiently well-separated.

\subsection{Underdetermined Systems, Coherence, and Sparse Solutions}
\label{subsec:sparsesols}

Let $\mathbf{d}\in\mathbb{R}^{\sensors}$ denote the noisy data vector with components given by (\ref{noisemodel1}).
Under the assumption that the number of sensors is much smaller than the number of potential emitter locations ($\sensors<<\emlocs$), the system $\bfPhi\bfp = \mathbf{d}$ to be solved for $\bfp$
 (an estimate of $\bfp_0$) is underdetermined, and so almost certainly possesses infinitely many solutions. However, as noted we will make the reasonable assumption that
there are few emitters, so that the solution vector $\bfp_0$ is sparse. More specifically, a vector $\bfp$ is said to be \textit{$k$-sparse} if $\bfp$ has at most $k$ nonzero components. Under the assumption that $\bfp_0$ is $k$-sparse for sufficiently small $k$, it is highly likely that a physically relevant solution can be found,
although the existence of a unique sparse solution and the ease with which it can be found depend on the measurement matrix $\bfPhi$.

One property that $\bfPhi$
can possess that leads to favorable recovery results is that of low ``mutual coherence.''
First, the \textit{coherence} of vectors $\bfx,\bfy\in\mathbb{R}^{\emlocs}$ is the quantity
\begin{equation}
\label{coher}
\mu(\bfx,\bfy) = \frac{|\bfx\cdot\bfy|}{\|\bfx\|_2\|\bfy\|_2}.
\end{equation}
The Cauchy-Schwarz inequality shows that $0\leq \mu(\bfx,\bfy)\leq 1$, with $\mu(\bfx,\bfy)=0$ when $\bfx$ and $\bfy$ are orthogonal and $\mu(\bfx,\bfy)=1$ when one vector is a scalar multiple of  another.
The \textit{mutual coherence} of an $\sensors\times \emlocs$ matrix $\bfPhi$ with columns $\bfPhi_i$ is the quantity
\begin{equation}
\label{matcoher}
\mu(\bfPhi) = \max_{i\neq j} \mu(\Phi_i,\Phi_j).
\end{equation}
Again, $0\leq \mu(\bfPhi)\leq 1$.  If $\mu(\bfPhi)=1$ then two or more distinct columns of $\bfPhi$ are scalar multiples of each other, while $\mu(\bfPhi)=0$ means $\bfPhi$ is
an orthogonal matrix, which is impossible in the present situation since $\sensors < \emlocs$.

Low coherence matrices are desirable when seeking sparse solutions to a linear system $\bfPhi\bfp=\mathbf{d}$. It can be shown that if $\mu(\bfPhi)<1/(2k-1)$ then any $k$-sparse solution
$\bfp$ is unique and many compressed sensing algorithms will converge to this solution (see Section 5.1 of \cite{Foucart}). Low mutual coherence also leads to more favorable bounds on the error in the presence of noisy data (\cite{donoho1}).

Unfortunately, for the localization problem described above, low mutual coherence will not hold for any realistic sensor configuration. First, our measurement matrix has entirely positive entries, so no cancelation occurs in the dot product of columns of $\bfPhi$; as a result, the pairwise coherence for any two columns is likely to be larger than for a matrix with mixed sign entries.  Also, if potential emitter locations $(x_i,y_i)$ and $(x_j,y_j)$ are closely spaced, then the $i$th and $j$th columns $\bfPhi_i$ and $\bfPhi_j$
of the measurement matrix will be nearly identical, and so have high pairwise coherence.  Thus, if we work on a fine grid (to obtain higher source resolution) we confront measurement matrices with high mutual coherence.
This presents a challenge for the finding the correct sparse solution.

\section{Algorithm for Sparse Solutions}
\label{sec:sparserecov}

In this section we briefly detail an algorithm appropriate for finding sparse solutions to the problem at hand. Our goal is not so much to focus on
this specific algorithm, but to use it to gain insight into the ill-posedness of this inverse problem, and provide examples that illustrate the analysis for resolution
and clearance.

\subsection{BLOOMP}
\label{subsec:bloomp}

Finding the sparsest solution to a linear system of equations is, in general, computationally intractable, even if a sparse solution is known to exist \cite{natarajan}.  However, a number of efficient computational approaches have been devised that, under the right conditions, find such a sparse solution with high probability.  In this section we justify use the algorithm ``Band-excluded Locally Optimized Orthogonal Matching Pursuit'' (BLOOMP, see \cite{FannjiangLiao}) for the present problem, and include an illustrative computational example.

Briefly, the BLOOMP algorithm is a modification of Orthogonal Matching Pursuit (OMP). OMP is a ``greedy'' algorithm that iteratively builds up a sparse solution to $\bfPhi\bfp=\mathbf{d}$ one nonzero component at a time. Let $\bfp^0=\mathbf{0}$ denote our initial guess at a solution, $\bfp^k$ the $k$th iterate (at most $k$-sparse) in OMP,
and $S^k=\{i: \bfp^k_i\neq 0\}$; $S^k$ is called the \textit{support} set of $\bfp^k$. The set $S^k$ indexes those columns of $\bfPhi$ that are being used to synthesize the data
$\mathbf{d}$. OMP constructs $\bfp^{k+1}$ by augmenting the support $S^k$ with a new index $i_k$ chosen so
that the residual $\|\bfPhi\bfp^{k+1}-\mathbf{d}\|_2$ is minimized. This continues until a maximum sparsity bound or a termination criterion is met.
One common stopping criterion takes the form $\|\bfPhi\bfp^k-\mathbf{d}\|_2 \leq C \epsilon$
where $\epsilon$ is comparable to the expected noise level in the data as measured in the Euclidean norm and $C\approx 1$; see \cite{Foucart}. We say more on this in our specific application below in Section \ref{subsec:example} and Appendix \ref{appA}.

A drawback of OMP is that once an index has been added to the support set $S^k$, it is never removed at a later iteration, so sub-optimal early choices cannot be
undone. Many modifications to OMP have been proposed to overcome this problem. We have adopted one such modification, BLOOMP \cite{FannjiangLiao}, because it is particularly suited to ``high-coherence'' measurement matrices.
Like OMP, BLOOMP builds a sparse solution by adding one index at each iteration to the potential support set. In our application this means adding one estimated emitter at each iteration.
However, in the BLOOMP algorithm
the column in $\bfPhi$ corresponding to the emitter added at a given iteration cannot have high coherence with any column of $\bfPhi$ corresponding to previously
added emitters. Physically, the next estimated emitter location cannot be too close to those already determined to be present---this is the ``band exclusion'' modification of OMP.
Moreover, at each iteration the emitters currently estimated to be present are subject to local adjustments in location and power to better fit the data; this is the
``local'' optimization portion of the algorithm. The authors in \cite{FannjiangLiao} show that in situations such as these---high coherence matrices, but in which the correct solution index support corresponds to columns with lower pairwise coherence, such as well-separated emitters on a finely-spaced grid---the BLOOMP modifications increase the probability of recovering the correct solution support indices, or in our case, the correct emitter number and location(s). We also add a constraint to the algorithm to require that at each iteration the emitter
power estimates must remain nonnegative.

\subsection{Recovery Example}\label{subsec:example}

\begin{figure}[ht]
\begin{center}
\includegraphics[width=\columnwidth]{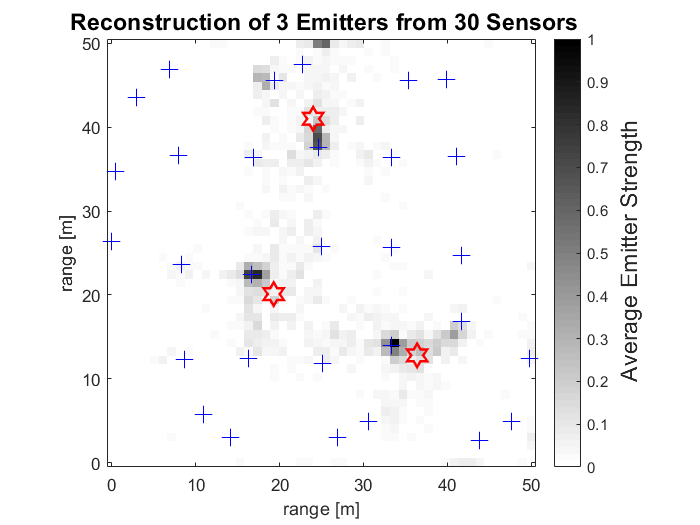}
\end{center}
\caption{Average recovered power from $500$ simulation runs (white is $0$ power, black is power $1$ or higher). The true emitters are marked as stars, sensor locations as crosses. The lognormal randomized noise is simulated with  $\sigma_{dB} = 3$.}\label{fig1}
\end{figure}

To illustrate, let $\Omega$ be the $50\times 50$ meter region $\{(x,y); 0\leq x,y\leq 50\}$ and consider a $50\times 50$ rectangular grid for potential emitter locations, of the form $(x_i,y_j)$ where $x_i=(i-0.5),y_j=(j-0.5)$ for $1\leq i,j\leq 50$, so here $\emlocs=50^2=2500$. In many settings it is the case that randomness in the construction of the measurement matrix is an asset in using sparsity or compressed sensing recovery algorithms \cite{candeswat}.
We thus consider $\sensors=30$ RSS data points collected from an emitter on a ``meandering'' path above $\Omega$, at altitude $h=10$ meters. The sensor locations are displayed as crosses in Fig. \ref{fig1}. We use pathloss exponent $\ploss=3.5$ in equation (\ref{attenmodel0}) (assumed known for now) and noise level
$\sigma_{dB}=3$ dB in equation (\ref{noisemodel1}). See Section \ref{sec:data} for data that supports these parameter choices, and for a recovery from measured data.

Three emitters with unit power at reference distance $r_0=1$ meter are placed at $(24.0,41.0)$, $(19.3,20.1)$, and $(36.4,12.8)$ (hereafter
referred to as emitters $1,2,$ and $3$). Note that
these are not themselves grid points; nonetheless, one would hope to recover emitter estimates that correspond to nearby grid points.
We then simulate noisy data $\mathbf{d}$ using equation (\ref{noisemodel1}) and perform a reconstruction from $\mathbf{d}$ using the BLOOMP algorithm, to recover an estimate
of the emitter number, location(s), and power(s). This process of generating noise and reconstructing is repeated $500$ times, each with a different noise realization.
The number of emitters is not assumed a priori.

One can show (see Appendix \ref{appA}) that for a modest noise level  $\sigma_{dB}\leq 5 \textrm{ dB}$ the expected value of $\|\mathbf{d}-\mathbf{d}_0\|^2_2$ is bounded by and comparable to the quantity $\epsilon = (\mu_0^2 + \sigma_0^2)\|\mathbf{d}_0\|_2^2$ where $\mu_0 = e^{\eta^2\sigma_{dB}^2/2}-1$ and $\sigma_0^2 = e^{\eta^2\sigma_{dB}^2}(e^{\eta^2\sigma_{dB}^2}-1)$ (recall $\eta = \ln(10)/10$). Of course we expect the noiseless data $\mathbf{d}_0$ is unknown, but the noisy data $\mathbf{d}$ provides a reasonable estimate. We thus terminate the iteration when the fit to the data is comparable to (or a bit smaller than) this noise level, specifically, when
\begin{equation}
\label{termcrit}
\|\mathbf{d}-\mathbf{d}_k\|_2 \leq C \sqrt{(\mu_0^2 + \sigma_0^2)}\|\mathbf{d}\|_2
\end{equation}
where $\mathbf{d}_k=\bfPhi\bfp^k$ denotes the estimated data at the $k$th iteration of BLOOMP and $C$ is a constant less than $1$ (we use $C=1/4$). For high noise
levels the random variable $\|\mathbf{d}-\mathbf{d}_k\|_2$ is more highly skewed to the right, and so $E(\|\mathbf{d}-\mathbf{d}_k\|_2)$ may be somewhat smaller than
$\sqrt{\|\mathbf{d}-\mathbf{d}_k\|_2^2}$, hence a value of $C$ somewhat less than $1$ can be helpful to prevent the iterative algorithm from terminating too early.

The results of these 500 simulated cases are shown in Fig. \ref{fig1}. The image is an average of the recovered power at each grid location, coded so $0$ recovered power is white, $1$ or higher is black. The sensor locations are illustrated as crosses and the true position of each emitter is represented by a star. The average estimated power for each of the three emitters is $1.7424, 0.8830,$ and $1.1712$ for emitters $1,2,$ and $3$.
The grey areas indicating positive power recovery clustered around the three emitters, which are reasonably well resolved. The spread of each cluster gives an indication
of the resolution one can achieve with this sensor configuration and noise level. An analysis of this resolution is the focus of the next section.

\begin{figure}[ht]
\begin{center}
\includegraphics[width=\columnwidth]{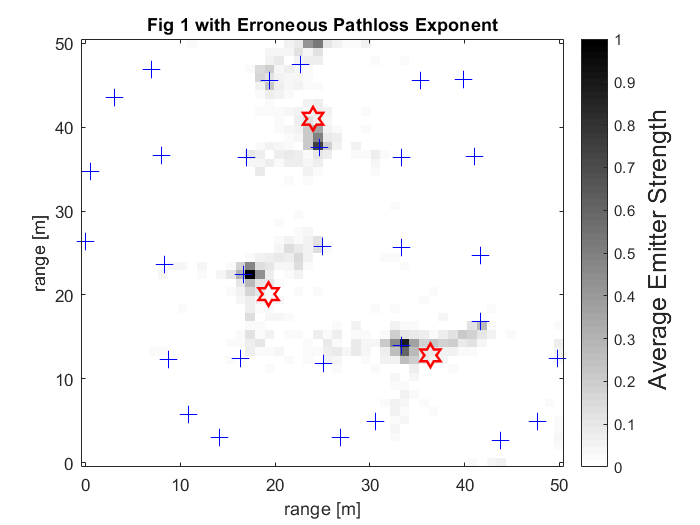}
\end{center}
\caption{Average recovered power from $500$ simulation runs (white is $0$ power, black is power $1$ or higher), erroneous pathloss exponent. The true emitters are marked as stars, sensor locations as crosses. The lognormal randomized noise is simulated with  $\sigma_{dB} = 3$.}\label{fig1b}
\end{figure}

The pathloss exponent $\ploss$ would appear to be a rather critical value in estimating the number and position of the emitters, but we find
that this is not the case. Specifically, an incorrect pathloss exponent has little effect on the recovery of the emitter count and locations, but does significantly affect the estimated power of each emitter. As an illustration, in Fig. \ref{fig1b} is shown a recovery with exactly the same
parameters as Fig. \ref{fig1}, but with the (erroneous) assumption of a pathloss exponent of $2.5$ (whereas $\ploss=3.5$ was used to generate the data). The average estimated power is $0.1368, 0.0595,$ and $0.0612$ for emitters $1,2,$ and $3$, respectively,
considerably off from the correct values of $1$ for each. Nonetheless, the number and location are quite accurate.

\section{Analysis of Resolution and Detection Limits}
\label{sec:analysis}

The goal in this section is to develop a method for quantifying the local resolution one can obtain at any fixed potential emitter location from RSS data for a given noise level and sensor configuration, and to
provide a bound on the weakest emitters that can be reliably detected.

\subsection{Resolution Analysis}

Suppose an emitter lies at one of two potential locations, say $\bfq_1 = (x_1,y_1)$ or $\bfq_2=(x_2,y_2)$. We collect noisy RSS data from $\sensors$ sensors.
The goal is to determine at which location the emitter actually lies, with sufficiently high probability
(to be specified). If this can be done we will say the two potential locations are ``resolvable.''

Let $\mathbf{d}_k\in\mathbb{R}^{\sensors}$ denote the noiseless RSS data we would collect from an emitter at location $\bfq_k$, where $k=1$ or $k=2$.
This data vector is assumed to obey the model (\ref{attenmodel0}), with a single nonzero power location.
For convenience we define normalized data vectors
\begin{equation}
\label{dispvecs}
\mathbf{b}_1=\frac{\mathbf{d}_1}{\|\mathbf{d}_1\|_2}\textrm{   and   }\mathbf{b}_2=\frac{\mathbf{d}_2}{\|\mathbf{d}_2\|_2}
\end{equation}
so $\|\mathbf{b}_k\|_2=1$ for $k=1,2$. Note that the reference power $p_k$ will not matter in either case.

Suppose we collect noisy data $\mathbf{d}\in\mathbb{R}^{\sensors}$ from the sensors, stemming from an emitter at location $\bfq_1$;
the components of $\mathbf{d}$ are given by (\ref{noisemodel1}) (with only a single nonzero summand).  The goal is to use $\mathbf{d}$ to correctly
assign the emitter to location $\mathbf{q}_1$, as opposed to $\mathbf{q}_2$. If we formulate this as a compressed sensing problem as above in equation (\ref{attenmodel}) then we obtain linear system
$$
\smtwo \mathbf{b}_1 & \mathbf{b}_2 \em \sv p_1\\p_2 \ev = \mathbf{d}
$$
(equivalently, $p_1\mathbf{b}_1 + p_2\mathbf{b}_2 = \mathbf{d}$)
in which the sensing matrix $\bfPhi$ is $\sensors\times 2$ with unit norm columns $\mathbf{b}_1$ and $\mathbf{b}_2$. We seek a $1$-sparse solution to this system.
In this very simple case OMP or any standard sparse solver (e.g. basis pursuit) will provide a $1$-sparse solution consisting of a multiple of that column of $\bfPhi$
which has the highest coherence with the data $\mathbf{d}$, with power estimate $\tilde{p}_k = \mathbf{b}_k\cdot\mathbf{d}$ for either $k=1$ or $k=2$. That is, the emitter
 is correctly assigned to location $\mathbf{b}_1$ if
\begin{equation}
\label{cohercomp}
\mu(\bfb_1,\mathbf{d}) > \mu(\bfb_2,\mathbf{d})
\end{equation}
and incorrectly to location $\mathbf{p}_2$ otherwise.
Condition (\ref{cohercomp}) is quite natural---the emitter is assigned to a location according to which vector $\mathbf{b}_1$ or $\mathbf{b}_2$ best matches the collected data $\mathbf{d}$ after optimal scaling for power. This notion of resolution is not wedded to a compressed sensing approach to the problem, nor any particular algorithm.

Equation (\ref{cohercomp}) is equivalent to
\begin{equation}
\label{cohercomp3}
\mathbf{c} \cdot \mathbf{d} > 0
\end{equation}
where
\begin{equation}
\label{cdef}
\mathbf{c} = \bfb_1-\bfb_2.
\end{equation}
Equations (\ref{cohercomp3}) and (\ref{cdef}) can be written equivalently as $Q>0$ where
\begin{equation}
\label{Qdef}
Q = \sum_{j=1}^{\sensors} w_j e^{\eta R_j}
\end{equation}
with $w_j=c_j/r_{1j}^{\ploss}$ and where $c_j$ denotes the $j$th component of $\mathbf{c}$. For a given sensor configuration the $w_j$ are known. We want to compute the probability $P(Q>0)$, so that we correctly assign the emitter to location $\bfq_1$. It should be noted that we will have $0.5\leq P(Q>0)\leq 1$, with $P(Q>0)=1$ as the best case---the
emitters are certainly resolvable---and $P(Q>0)=0.5$ as the worst case, in which resolving the emitter locations becomes a ``coin toss.''

The random variable $Q$ is a signed linear combination of lognormal random variables (the $w_i$ are generally of mixed sign).
The next section is devoted to accurately approximating the probability $P(Q>0)$ in an easily computable fashion.

\subsection{Approximating a Signed Sum of Lognormal Random Variables}

Though $Q$ has coefficients of mixed sign, we first consider the case in which all coefficients are positive.
The distribution of such a sum of lognormals is a well-studied problem, though such a sum has no closed-form density function.
However, it has long been noted that such a sum is itself approximately lognormal, and so can be characterized as being of the form $e^{N(\mu,\sigma^2)}$ for
suitable $\mu$ and $\sigma$ (Note that $\mu$ stands for the mean of the noise distribution here, not mutual coherence).

In \cite{mehta} the authors provide a simple and effective method for fitting $\mu$ and $\sigma$ to such a sum. The individual lognormals in the sums
they consider are of the form $e^{N(\mu_i,\sigma_i^2)}$
with varying $\mu_i$ and $\sigma_i$, and are assumed independent. For a linear combination of the form (\ref{Qdef}) with weights $w_i$ that are positive, the weighted sum in $Q$ is easily adapted to this setting, by absorbing
the $w_i$ into the $R_i$ (we can shift the mean of $R_i$ by $\ln(w_i)$). If we split the sum defining $Q$ into a piece with positive weights and a piece with negative weights, we can write $Q = Q^+ - Q^-$ where
\begin{equation}
\label{Qdefposneg}
Q^+ = \sum_{w_j\geq 0} w_j e^{\eta R_j}\textrm{  and  }Q^- = \sum_{w_j<0} (-w_j) e^{\eta R_j}.
\end{equation}
The method of \cite{mehta} provides a lognormal random variable approximation for $Q^+$ in the form $e^{R}$ where $R=N(\mu_+,\sigma_+)$, by determining an appropriate mean and variance $\mu_+$ and $\sigma_+^2$.  A similar approximation is made to obtain $\mu_-$ and $\sigma_-$ for $Q^-$.

The probability density function (pdf) and cumulative density function (cdf) for the lognormal random variable are well-known.
Moreover, if a random variable $X$ has cdf $F(x)$ and random variable $Y$ has pdf $g(x)$ then the cumulative distribution function $H(x)$ of $X-Y$ is given by
$$
H(x) = \int_{-\infty}^{\infty} F(x+y) g(y)\,dy.
$$
Then, for example, $P(X-Y>0)$ is given by $1-H(0)$.
In the present case the cdf $H$ for $Q=Q^+-Q^-$ can be expressed as
\begin{align}
H(x) & = \int_{\max(0,-x)}^{\infty} \left [\frac{1}{2}+\frac{1}{2}{\rm erf}\left (\frac{\ln(x+y)-\mu_+}{\sigma_+\sqrt{2}} \right )\right ]\nonumber\\
& \times \left [ \frac{1}{y\sigma_-\sqrt{2\pi}} e^{-\frac{(\ln(y)-\mu_-)^2}{2\sigma_-^2}}\right ]\,dy.\label{Qcdf}
\end{align}
The $\max(0,-x)$ lower limit cuts off the integral as soon as the cdf or pdf of either random variable equals zero. The value we are interested in is $P(Q>0)=1-H(0)$,
and this can be computed easily from (\ref{Qcdf}).

The overall procedure is as follows: Given potential emitter locations $\bfq_1$ and $\bfq_2$, we compute $\mathbf{c}$ as
in (\ref{cdef}) and set $w_j=c_j/r_{1j}^{\ploss}$ with $r_{1j}$ as the distance from location $\bfq_1$ to
the $j$th sensor. We then use the procedure in \cite{mehta}) to estimate $\mu_+,\sigma_+,\mu_-,$ and $\sigma_-$ for $Q^+$ and $Q^-$ and compute $P(Q>0)$ using (\ref{Qcdf}).
If $P(Q>0)$ exceeds some threshold probability $p_{min}$ we will say the emitter location $\bfq_1$ is resolvable from location $\bfq_2$.

\begin{figure}[ht]
\begin{center}
\includegraphics[width=\columnwidth]{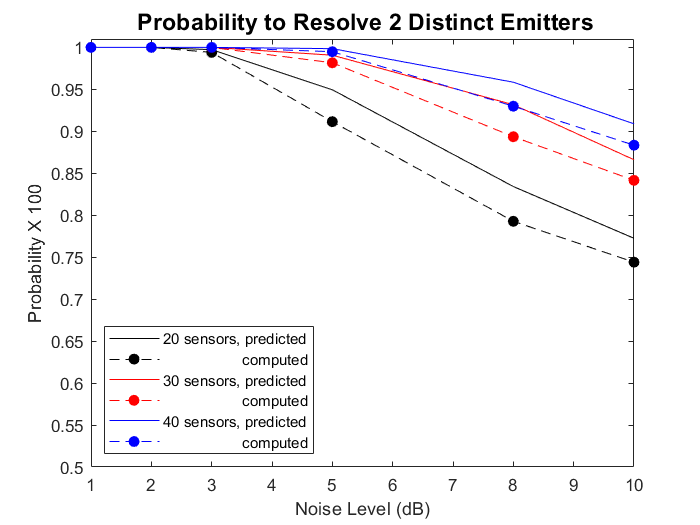}
\end{center}
\caption{Simulated and approximated resolution probability for various sensor counts and noise levels.}
 \label{probnoisefig}
\end{figure}

To illustrate the accuracy of the approximation, Fig. \ref{probnoisefig} shows the quantity $P(Q>0)$ computed by this procedure versus the simulated probability of correctly resolving the emitters locations for a variety of sensor
counts and noise levels. In each base we use $\bfq_1=(24.5,41.5)$ and $\bfq_2=(19.5,20.5)$ with sensors at random $(x,y)$ locations in $0<x,y<50$ at altitude $h=10$. We generate $10^4$ realizations of synthetic noisy data $\mathbf{d}$ for a sensor at location $\bfq_1$ and assign it to location $\bfq_1$ if $\mu(\mathbf{d},\mathbf{b}_1)>\mu(\mathbf{d},\mathbf{b}_2)$, location $\bfq_2$ otherwise. The pathloss exponent is $3.5$.

\begin{figure}[ht]
\begin{center}
\includegraphics[width=\columnwidth]{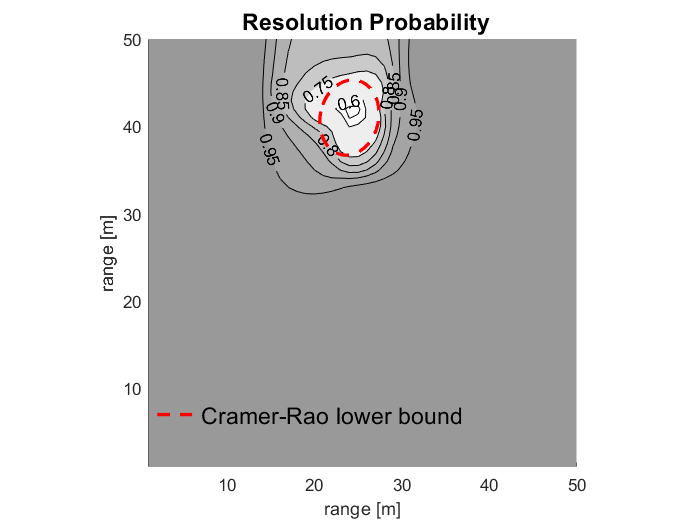}
\end{center}
\caption{Probability of successful resolution as function of $(x,y)$. Red oval is a
95 percent confidence region from the Cramer-Rao bounds.}
 \label{resfig1}
\end{figure}

As an example of how this can be used to quantify local resolution, consider the three-emitter configuration of Fig. \ref{fig1}, with the same noise level and other parameters. What local resolution might we expect near the emitter at location $(24,41)$? Let $\mathbf{p}_1=(24,41)$ and $\mathbf{p}_2 = (x,y)$ for $0<x,y<50$, so that $P(Q>0)$ as computed above is a function of $(x,y)$. In Fig. \ref{resfig1} we show a contour plot of this function.
The red oval delineates, for comparison, the Cramer-Rao lower bounds on the uncertainty in estimating the location of the emitter (discussed below).

\begin{figure}[ht]
\begin{center}
\includegraphics[width=\columnwidth]{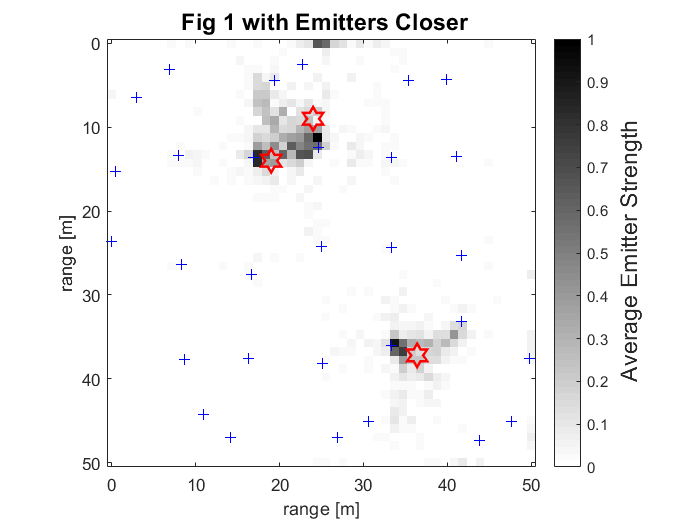}
\end{center}
\caption{Average recovered power from $500$ simulation runs (white is $0$ power, black is power $1$ or higher). The true emitters are marked as stars, sensor locations as crosses.}\label{fig1h}
\end{figure}

To illustrate the validity of the resolution analysis, in Fig. \ref{fig1h} is shown a situation similar to that of Fig. \ref{fig1}, but in which the emitter
at position $(19.3,20.1)$ has been moved to $(19.0,36.0)$, which is only 7 meters away from the emitter at position at $(24,41)$. The newly moved emitter lies outside the Cramer-Rao bounds, on about the $P(Q>0)=0.85$
contour. The emitters are not as reliably resolved.

This analysis makes it clear that, for a given noise level (and other parameters) the resolution obtainable with RSS data is limited,
and can be quantified. In particular, in a compressed
sensing approach there are little improvements in resolution to be obtained by using too fine of a grid.

\subsection{Comparison to Cramer-Rao Bounds}

Other authors (e.g., \cite{Chitte1,Chitte2,Koorapaty}) have examined statistical bounds, for example, Cramer-Rao bounds, on the minimum variance that can obtained by using RSS
data to estimate the distance to or position of an emitter. Such a bound provides a natural way to quantify resolution. However, as noted in \cite{Chitte1,Chitte2},
the Cramer-Rao bounds in this setting cannot be attained by any unbiased estimator, and so are too optimistic. (It should also be noted that our estimates are almost
certainly biased.)

To illustrate and compare with the current
analysis, we consider a single emitter of unknown power $p_0$ at true location $(22,41)$, altitude zero, with the $30$ sensor locations as used in Figs. \ref{fig1}, \ref{resfig1}, and \ref{fig1b}, pathloss exponent $3.5$, and noise level $3$ dB. Following the computations of Section 3.2 in \cite{Chitte2} we establish a Cramer-Rao lower bound on the minimum covariance of any unbiased estimator of the emitter location and power. The red elliptical region in Fig. \ref{resfig1} is a 95 percent confidence region with respect to the spatial variables for
an emitter with $p_0=1$, though the bounds do not depend on the unknown power $p_0$. The lower bound on the variance of any unbiased estimate of
$p_0$ is $0.16$.

\subsection{Detectability and Clearance}

In this section we consider the problem of when we can be reasonably certain that we have detected all the emitters above a given power threshold in a region of interest; this
could be the entire region $\Omega$ or some subregion thereof.

For a given configuration of $\sensors$ sensors, $\emlocs$ potential emitter locations and corresponding $\sensors\times \emlocs$ measurement matrix $\bfPhi$, suppose that $\bfp_0\in\mathbb{R}^{\emlocs}$ embodies the true emitter power vector. The noise-free data $\mathbf{d}_0\in\mathbb{R}^{\sensors}$ is given by (\ref{attenmodel}); let $\mathbf{d}\in\mathbb{R}^{\sensors}$ be the collected (noisy) data vector.
Suppose that $\bfp_r$ is an estimate of $\bfp_0$ based on the data $\mathbf{d}$, computed using BLOOMP or any other recovery algorithm.  We assume, however, that the algorithm produces an estimate $\bfp_r$ for which an error bound of the form $\|\bfPhi\bfp_r-\mathbf{d}\|\leq \epsilon$ holds, for some tolerance $\epsilon$, where $\|\cdot\|$ can denote any norm, e.g., the $L^2$ or supremum norm. Typically $\epsilon$ is comparable to the expected noise level in the data in the appropriate norm.

Now suppose that a single additional emitter were present at location $\bfr_i$, with power $P$. Let $\tilde{\bfp}=\bfp_r+P\bfe_i$ denote resulting power vector ($\bfe_i$ is the $i$th standard basis vector). This would yield data
$
\tilde{\mathbf{d}} = \bfPhi\tilde{\bfp} = \bfPhi\bfp_r + P \bfPhi_i.
$
We will consider the additional emitter at $\bfr_i$ to be detectible if
\begin{equation}
\label{detectable}
\|\tilde{\mathbf{d}}-\mathbf{d}\|>\epsilon.
\end{equation}
That is, the presence of this additional emitter would yield reconstructed data $\tilde{\mathbf{d}}$ that is inconsistent with the measured data at the given tolerance level. But we do not require that the reconstructed emitter power configuration $\bfp_r$ be accurate, in that  $\|\bfp_r-\bfp_0\|$ need not be small.

The value of $P$ that assures $\|\tilde{\mathbf{d}}-\mathbf{d}\|>\epsilon$ holds can be estimated. We have, using the reverse triangle inequality
\begin{eqnarray*}
\|\tilde{\mathbf{d}}-\mathbf{d}\| & = & \|P\bfPhi_i + \bfPhi\bfp_r-\mathbf{d}\|\nonumber\\
& \geq & \left |P \|\bfPhi_i\| - \|\bfPhi\bfp_r-\mathbf{d}\|\right |\nonumber\\
& \geq & P\|\bfPhi_i\| - \epsilon.
\end{eqnarray*}
Inequality (\ref{detectable}) must hold if $P\|\bfPhi_i\| - \epsilon>\epsilon$ or
\begin{equation}
\label{powthresh}
P>\frac{2\epsilon}{\|\bfPhi_i\|}.
\end{equation}
The threshold on the right in (\ref{powthresh}) depends on the precision to which we fit the measured data, i.e., the noise level in the data, the norm we use, and on $\bfPhi$. By taking the maximum of the right side of (\ref{powthresh}) over all locations $\bfr_i$ in a given region
$\Omega'\subseteq \Omega$ we obtain a threshold of the weakest emitters that can be reliably identified in $\Omega'$. If a lower
threshold is desired, it would be necessary to alter the number and/or placement of sensors. Inequality (\ref{powthresh}) quantifies what
is required. Of course the estimates leading to (\ref{powthresh}) are likely pessimistic---an emitter may well be detected below this power threshold---but it does provide a rough lower bound for emitter detectability.

\begin{figure}[ht]
\begin{center}
\includegraphics[width=\columnwidth]{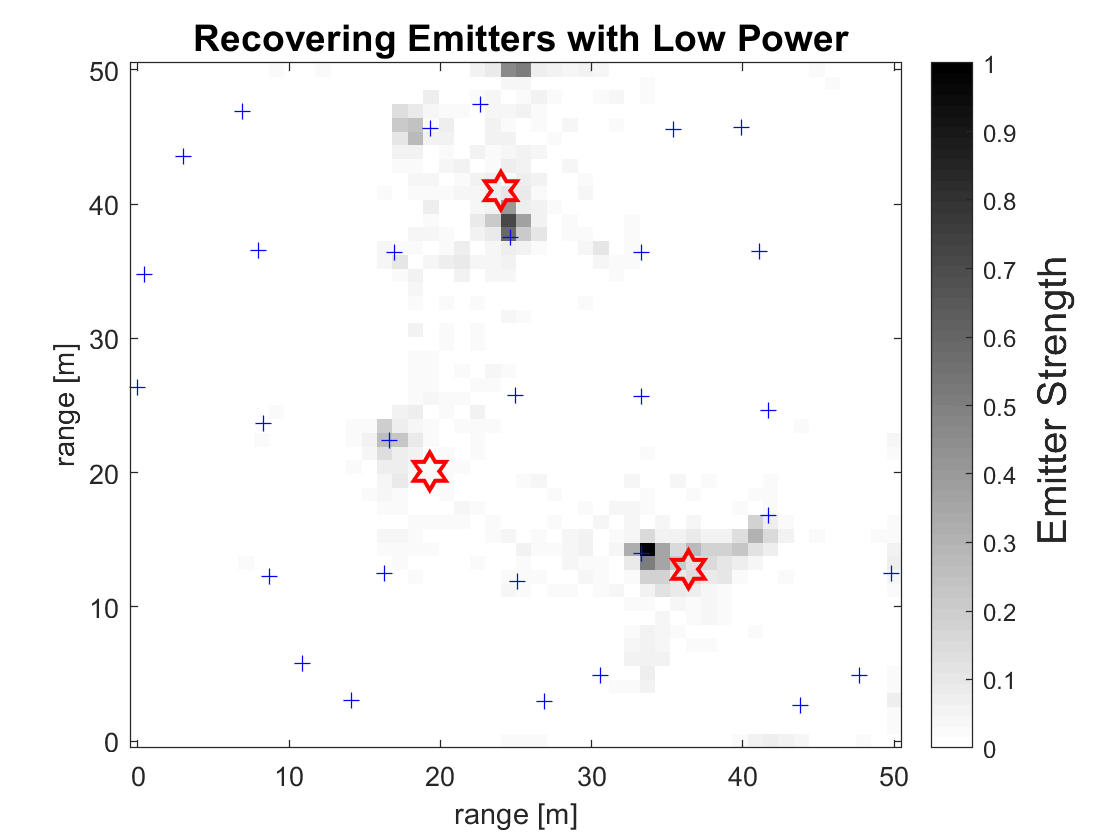}
\end{center}
\caption{Setting of Fig. \ref{fig1}, but with emitter at $(19.3,20.1)$ at power $0.5$.}
 \label{clear1}
\end{figure}

To illustrate, again consider the setting of Fig. \ref{fig1}. Let us consider the power threshold for detectability of the emitter at location $(19.3,20.1)$. The closest
grid location is $\bfr_{1020} = (19.5,20.5)$ (that is, index location $i=1020$ in our indexing scheme). We iterate BLOOMP until $\|\tilde{\mathbf{d}}-\mathbf{d}\|_2\leq 2\times 10^{-4}$
and compute $\|\bfPhi_{1020}\|_2 \approx 3.56\times 10^{-4}$, leading to a power bound $P\approx 1.1$ for the emitter in this location. As is obvious in Fig. \ref{fig1}, the emitter
is clearly detectable at power level $1$. However, under the same conditions but with power level $0.5$ the result is as shown in Fig. \ref{clear1}. At power level $0.25$
the emitter becomes essentially invisible.

\section{Illustration With Measured Data}
\label{sec:data}

In this section we briefly detail an experiment we, with the aid of our students, performed to collect actual RSS data under the conditions that were only simulated above. Our goal here is not
to reproduce the resolution or clearance analysis with experimentation, but rather to estimate realistic noise and pathloss parameters under relatively ideal conditions. However, we do perform a reconstruction for a single emitter, and illustrate the effect of using an erroneous pathloss exponent.

\subsection{Measurement of RSS in Open Air}
\label{subsec:rssmeas}

\begin{figure}[ht]
\begin{center}
\includegraphics[width=\columnwidth]{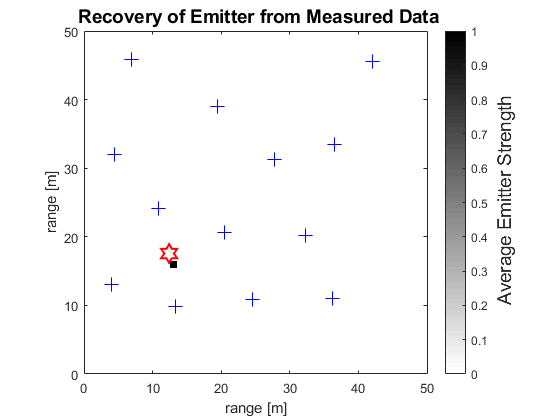}
\end{center}
\caption{The normalized RSS measured from sensors randomly placed in a 50x50 meter search area are plotted (dots). The log-linear fit is displayed as the dashed line. The pathloss trend predicted by the free-space approximation is displayed as a solid black line.}
 \label{reconfig2}
\end{figure}

An experiment to collect RSS data from a single emitter using $15$ sensors was conducted in the open-air on a flat grass-covered field of $90\times 120$ meters with no overhead obstructions. The transmitter was placed at location $(12.4,17.5)$ meters relative to an origin on a cartesian grid, at a height of $70$ cm. Fifteen different receivers were scattered within a $50\times 50$ meter square area to collect RSS samples at 15 different positions $\bfs_j$, corresponding to 15 different distances $r_{1j}$. The height of the receivers was $50$ cm. The locations of the receivers and emitter are plotted in Fig. \ref{reconfig2}. This isn't precisely the configuration we simulated, but the analysis is easily adapted to
any emitter/sensor geometry.

The transmitter emitted a continuous-wave, unmodulated signal, centered at 925 MHz (in the ISM band) using a Software Defined Radio (SDR) transceiver (USRP E100, Ettus Research). An omnidirectional vertical dipole antenna was used for the transmitter (VERT900, Ettus Research). The transmitted signal was sampled at each sensor position at a rate of $1.152$ Msamples/s, for a duration of one second, using an SDR radio (receiver only) device with a USB interface (R820T NESDR Mini, Noo Electric). The RTL-SDR has the capability to tune over the range 25 MHz to 1.75 GHz, producing raw, 8-bit IQ data samples, at a programable, baseband sampling rate of up to 2.8MHz \cite{stewart}. However, the data acquisition sampling rate was set lower to ensure the accuracy of the rate. The gain was set to $32.8$ for each of the receivers, which was tuned so that the receiver closest to the transmitter ($\approx 6.5$ meters away) did not experience saturation. Without automatic gain control, we found the useful dynamic range of the RTL-SDR is around $45$ dB. The receivers used an omnidirectional vertical dipole antenna, approximately $14$ cm in length with an MCX connection.

\begin{figure}[ht]
\begin{center}
\includegraphics[width=\columnwidth]{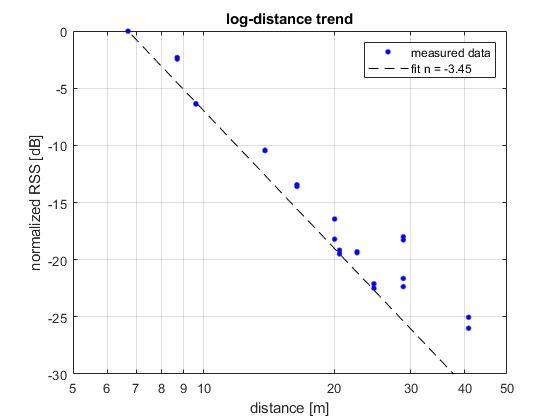}
\end{center}
\caption{The emitter was placed at the position (12.41m, 17.56m) as indicated by the red asterisk. The receiver locations are marked by crosses. An estimate of the emitter's location as computed by the BLOOMP algorithm is at the position (12.50,15.50) and the magenta colored pixel where the normalized power is estimated to be 100\%. }
 \label{reconfig1}
\end{figure}

The raw IQ data were processed using the procedures recommended in \cite{zanella}.  The RSS was calculated by first
applying a Chebyshev Type I IIR filter of order $1$ to remove most of the fast-fading variations The RSS values are normalized to the value received at the sensor with the shortest distance to the transmitter (about six meters). A least-squares fit to the log-normal distance trend is used to estimate the path-length exponent, $\ploss\approx 3.45$ for our data. The standard deviation of the log normalized uncertainty term (long-term fading uncertainty) was computed from the variation from the fitted data, $\sigma_{dB} \approx 1.86$ dB. The decimated RSS values and fit are plotted in Figure \ref{reconfig1}.

\subsection{A Sample Reconstruction from Data}
\label{subsec:datarecon}

A reconstruction is performed using the BLOOMP algorithm to reconstruct an estimate of the transmitter power vector, $\mathbf{p}$, from the measured RSS vector, $\mathbf{d}$. The index value of the power vector, $p_i$, which are not estimated to be zero will indicate the location of a detected emitter. Although our measured data is known to be received from only one emitter, we allow the algorithm to iterate as many as 12 times, corresponding to a reconstruction with the potential to predict as many as 12 emitter locations. The measurement matrix,  $\bfPhi$, is calculated by assuming a one-slope propagation model (\cite{zanella} and our equation (3)), with the reference distance set to 1, the path-loss coefficient set to 3.45, the received power at the reference distance, $K$, is set 1 and all of the RSS values are normalized to this reference power. Band exclusion is applied to the modified BLOOMP algorithm according to \cite{FannjiangLiao} with a exclusion parameter set to 0.98. The BLOOMP reconstruction algorithm is terminated after one iteration when the residuals are less than $0.5$ times the expected uncertainty using a log normalized standard deviation of 2dB. In Fig 2, the estimated location of the emitter is plotted as the magenta square. The estimate falls on the closest grid point at (12.50m,15.50m) which is 2.06 meters away from the position of the true emitter's location (ro = 12.41m, 17.56). It should be noted that the true emitter is not placed on a grid point.

\section{Conclusion}\label{sec:conclusion}

We have formulated the problem of geolocating multiple non-cooperative RF emitters in a given region using low-capability sensors as a problem in compressed sensing, and use this formulation to develop methods for examining the limits on resolution and emitter detectability as a function of
the data noise level, sensor number and configuration, as well as other relevant variables, for example, the pathloss exponent. We have also implemented an algorithm suitable for actually recovering emitter number and location from simulated data. We also demonstrate
the recovery of a single emitter using measured data.

Several natural extensions and refinements of this technique suggest themselves.  The model can be easily adapted to directional sensor antennae, and sensors (or emitters) at nonconstant altitude.  Also of interest, but more challenging, is the problem of locating anisotropic, intermittent, or moving emitters, and operating in an environment in which sensor positions' themselves are not known and must be estimated.

\section{Acknowledgements}\label{acknowledgements}

The authors would like to thank the many Rose-Hulman undergraduate and masters students who have worked with us on this project (and continue to do so).  Dr. Walter
would also like to thank the many colleagues at the Air Force Research Laboratory Sensors Directorate who have supported her.

\section{Appendix A}
\label{appA}

As noted in Section \ref{subsec:example}, we iterate the BLOOMP algorithm until the fit squared residual is comparable to
$E(\|\mathbf{d}-\mathbf{d_0}\|_2^2)$. The latter quantity can be estimated from $\mathbf{d}$ (the measured data) and the noise level $\sigma_{dB}$.

From the noise-free model (\ref{attenmodel0}) and (\ref{noisemodel1}) we compute
\begin{align}
\|\mathbf{d}-\mathbf{d}_0\|_2^2 & = d_0^{2\ploss} \sum_{i=1}^{\sensors} \left (\sum_{j=1}^{\emlocs} \frac{p_j}{r_{i,j}^{\ploss}}X_{i,j}\right )^2
\end{align}
where $X_{i,j} = e^{\eta R_{i,j}/10}-1$ with $R_{i,j}$ normal with mean zero, variance $\sigma^2_{dB}$. The random variable
$X_{i,j}$ is lognormal with mean and variance given by
\begin{equation}
\label{meanvar}
\mu_0 = e^{\eta^2\sigma_{dB}^2/2}-1,\;\;\;\sigma_0^2 = e^{\eta^2\sigma_{dB}^2}(e^{\eta^2\sigma_{dB}^2}-1).
\end{equation}
Since the expected value is linear,
\begin{align}
E(\|\mathbf{d}-\mathbf{d}_0\|_2^2) & = d_0^{2\ploss} \sum_{i=1}^{\sensors} E\left [ \left (\sum_{j=1}^{\emlocs} \frac{p_j}{r_{i,j}^{\ploss}}X_{i,j}\right )^2\right ]
\end{align}
A little algebra shows that
\begin{align}
E \left [\left (\sum_{j=1}^{\emlocs} \frac{p_j}{r_{i,j}^{\ploss}}X_{i,j}\right )^2\right ] & = E \left [\sum_{j,k=1}^{\emlocs} \frac{p_jp_k}{r_{i,j}^{\ploss} r_{i,k}^{\ploss}} X_{i,j}X_{i,k}\right ]\nonumber\\
& = \sum_{j,k=1}^{\emlocs} \frac{p_jp_k}{r_{i,j}^{\ploss} r_{i,k}^{\ploss}} E(X_{i,j}X_{i,k})\nonumber\\
& = \sum_{j=1}^{\emlocs} \frac{p_j^2}{r_{i,j}^{2\ploss}} E(X_{i,j}^2)\nonumber\\
& + \sum_{j,k=1,j\neq k}^{\emlocs} \frac{p_jp_k}{r_{i,j}^{\ploss} r_{i,k}^{\ploss}} E(X_{i,j}X_{i,k})
\end{align}
Since the $X_{i,j}$ are independent we have
\begin{align}
E(X_{i,j}X_{i,k}) & = E(X_{i,j})E(X_{i,k}) = \mu_0^2\nonumber\\
E(X_{i,j}^2) & = \mu_0^2 + \sigma_0^2. \nonumber
\end{align}
Then some mundane algebra shows that
\begin{align}
& E(\|\mathbf{d}-\mathbf{d}_0\|_2^2)\nonumber\\
& = d_0^{2\ploss} \sum_{i=1}^{\sensors} \left ((\mu_0^2 + \sigma_0^2)  \sum_{j=1}^{\emlocs} \frac{p_j^2}{r_{i,j}^{2\ploss}}\right . \nonumber\\
& + \left . \mu_0^2 \sum_{j,k=1,j\neq k}^{\emlocs} \frac{p_jp_k}{r_{i,j}^{\ploss} r_{i,k}^{\ploss}}\right )\nonumber\\
& = d_0^{2\ploss} \sum_{i=1}^{\sensors} \left (\mu_0^2 \sum_{j,k=1}^{\emlocs} \frac{p_jp_k}{r_{i,j}^{n}r_{i,k}^{\ploss}} + \sigma_0^2 \sum_{j=1}^{\emlocs} \frac{p_j^2}{r_{i,j}^{2\ploss}}\right )\nonumber\\
& = d_0^{2\ploss} \sum_{i=1}^{\sensors} \left (\mu_0^2 \left (\sum_{j=1}^{\emlocs} \frac{p_j}{r_{i,j}^{n}}\right )^2 + \sigma_0^2 \sum_{j=1}^{\emlocs} \frac{p_j^2}{r_{i,j}^{2\ploss}}\right )\nonumber\\
& = \mu_0^2 \|\mathbf{d}_0\|_2^2 + \sigma_0^2 d_0^{2\ploss} \sum_{i=1}^{\sensors} \sum_{j=1}^{\emlocs} \frac{p_j^2}{r_{i,j}^{2\ploss}}\nonumber\\
& \leq \mu_0^2 \|\mathbf{d}_0\|_2^2 + \sigma_0^2 d_0^{2\ploss} \sum_{i=1}^{\sensors} \sum_{j,k=1}^{\emlocs} \frac{p_jp_k}{r_{i,j}^{\ploss} r_{i,k}^{\ploss}}\nonumber\\
& = \mu_0^2 \|\mathbf{d}_0\|_2^2 + \sigma_0^2 \sum_{i=1}^{\sensors} \left (d_0^{\ploss} \sum_{j}^{\emlocs} \frac{p_j}{r_{i,j}^{\ploss}}\right )^2\nonumber\\
& = \mu_0^2 \|\mathbf{d}_0\|_2^2 + \sigma_0^2\|\mathbf{d}_0\|_2^2\nonumber\\
& = (\mu_0^2 + \sigma_0^2)\|\mathbf{d}_0\|_2^2.\label{expresid}
\end{align}
This provides the basis for the termination criterion (\ref{termcrit}) (replacing $\mathbf{d}_0$ with $\mathbf{d}$.)

Note however that $E(\|\mathbf{d}-\mathbf{d}_0\|_2^2) \geq E(\|\mathbf{d}-\mathbf{d}_0\|_2)^2$, so the termination criterion (\ref{termcrit}) may result in under-fitting the data.
When $\sigma_{dB}\leq 5$ the quantities $E(\|\mathbf{d}-\mathbf{d}_0\|_2^2)$ and $E(\|\mathbf{d}-\mathbf{d}_0\|_2)^2$ are comparable in magnitude, but for larger noise
levels the random variable $\|\mathbf{d}-\mathbf{d}_0\|_2$ is skewed heavily higher, to the right. In such a case a smaller value of $C$ in (\ref{termcrit}) is appropriate.

\end{document}